\documentclass{article}
\usepackage[utf8]{inputenc}
\usepackage[margin=1in, includehead, includefoot]{geometry}
\usepackage{amsthm,amssymb,amsmath}
\usepackage{enumitem}
\usepackage{mathrsfs}
\usepackage{mathtools}
\usepackage{fancyhdr}
\usepackage{indentfirst}
\usepackage{graphicx}
\usepackage{subfigure}
\usepackage{placeins}
\usepackage{wrapfig}
\usepackage{url}
\usepackage{array}
\usepackage[numbers]{natbib}
\usepackage[hidelinks]{hyperref}

\newtheorem{definition}{Definition}
\newtheorem{lemma}{Lemma}[section]
\newtheorem{theorem}{Theorem}[section]
\newtheorem{corollary}{Corollary}[section]
\newtheorem{observation}{Observation}[section]

\newcommand{\toroman}[1]{\textit{\expandafter{\romannumeral #1\relax}}}

\newcommand{\npcompleteproblem}[3]{ \par\vspace{0.5em}\noindent{\textbf{#1}}\newline \textbf{INSTANCE: } #2 \newline \textbf{QUESTION: } #3 \par\vspace{0.5em} }

\newcommand{\centered}[1]{\begin{tabular}{@{}c@{}} #1 \end{tabular}}

\newcommand{\bisharp}[1]{\overset{\text{\large$_{_{\overleftrightarrow{}}}$}}{#1}}
\renewcommand{\sharp}[1]{{#1}^\#}

\title{Progress on Self Identifying Codes}
\author{
    \small Devin Jean \\
    \small Computer Science Department \\
    \small Middle Tennessee State University \\
    \small \texttt{devin.jean@mtsu.edu}
    \and
    \small Suk Seo \\
    \small Computer Science Department \\
    \small Middle Tennessee State University \\
    \small \texttt{Suk.Seo@mtsu.edu}
}
\date{}

\begin{document}
\maketitle
\thispagestyle{empty}

\begin{abstract}
The concept of an identifying code for a graph was introduced by Karpovsky, Chakrabarty, and Levitin in 1998 as the problem of covering the vertices of a graph such that we can uniquely identify any vertex in the graph by examining the vertices that cover it.
An application of an identifying code would be to detect a faulty processor in a multiprocessor system.
In 2020, a variation of identify code called ``self-identifying code'' was introduced by Junnila and Laihonen, which simplifies the task of locating the malfunctioning processor.
In this paper, we continue to explore self-identifying codes.
In particular, we prove the problem of determining the minimum cardinality of a self-identifying code for an arbitrary graph is NP-complete and we investigate minimum-sized self-identifying code in several classes of graphs, including cubic graphs and infinite grids.

\end{abstract}


\section{Introduction}
Let $G = (V(G), E(G))$ be a graph; we will assume all graphs are simple, undirected, and connected.
If $G$ is an infinite graph, we will assume it has the ``slow-growth'' property \cite{sampaio2024density}, which enables the use of many traditional finite density proof techniques.
The \emph{open neighborhood} of a vertex $v \in V(G)$, denoted $N(v)$ is the set of vertices adjacent to $v$; that is, $N(v) = \{ u \in V(G) : uv \in E(G) \}$.
The \emph{closed neighborhood} of a vertex $v \in V(G)$, denoted $N[v]$ is the set of vertices adjacent to $v$, plus $v$ itself; that is, $N[v] = N(v) \cup \{v\}$.

A \emph{dominating set} for $G$ is a subset of vertices $S \subseteq V(G)$ such that every vertex in the graph is within distance one of some vertex in $S$, or more formally, $\cup_{v \in S} N[v] = V(G)$.
An \emph{identifying code} (IC) for $G$ is a special subcategory of dominating sets which have the additional property that for any distinct $u,v \in V(G)$, $N[v] \cap S \neq N[u] \cap S$; for brevity, we will use the shorthand $N_S[v] = N[v] \cap S$, which is also known as the ``locating code'' of $v$.
Often, we interpret the vertices in $S$ as locations equipped with some form of ``detector'' or ``sensor'' which can transmit either 0 or 1 based on the absence or presence of an ``intruder'' in their closed neighborhood.
In this interpretation, the uniqueness requirement of closed neighborhoods intersected with $S$ gives rise to a distinct pattern of detector ``alarms'' for each vertex, thus allowing us to precisely locate the intruder location in the graph.

Suppose $S \subseteq V(G)$ is an identifying code for $G$.
If there is an intruder at some vertex $x \in V(G)$, we will observe a subset of detector vertices $A \subseteq S$ which sensed an intruder in their closed neighborhoods; specifically, we know that $A = N_S[x]$.
In practice, we are given only the set $A$ and must determine the intruder location, $x$.
We know that $x \in B$ where $B = \cap_{v \in A} N[v]$.
To precisely locate the intruder, we begin with this set of candidate locations, $B$, and perform a process of elimination based on the differences in locating codes of vertices in $B$.
For example, if we have two candidate intruder locations, $a,b \in B$, then we know there must be some detector vertex $w \in N_S[a] \triangle N_S[b]$.
Therefore, we know $w \notin A$, which implies that $x \notin N[w]$; this eliminates exactly one of $a$ or $b$, and possibly other unrelated candidates.
This ability to always eliminate one of a pair of candidate intruder locations is also known as being able to ``distinguish'' said pair of vertices.

\begin{figure}[ht]
    \centering
    \includegraphics[width=0.175\linewidth]{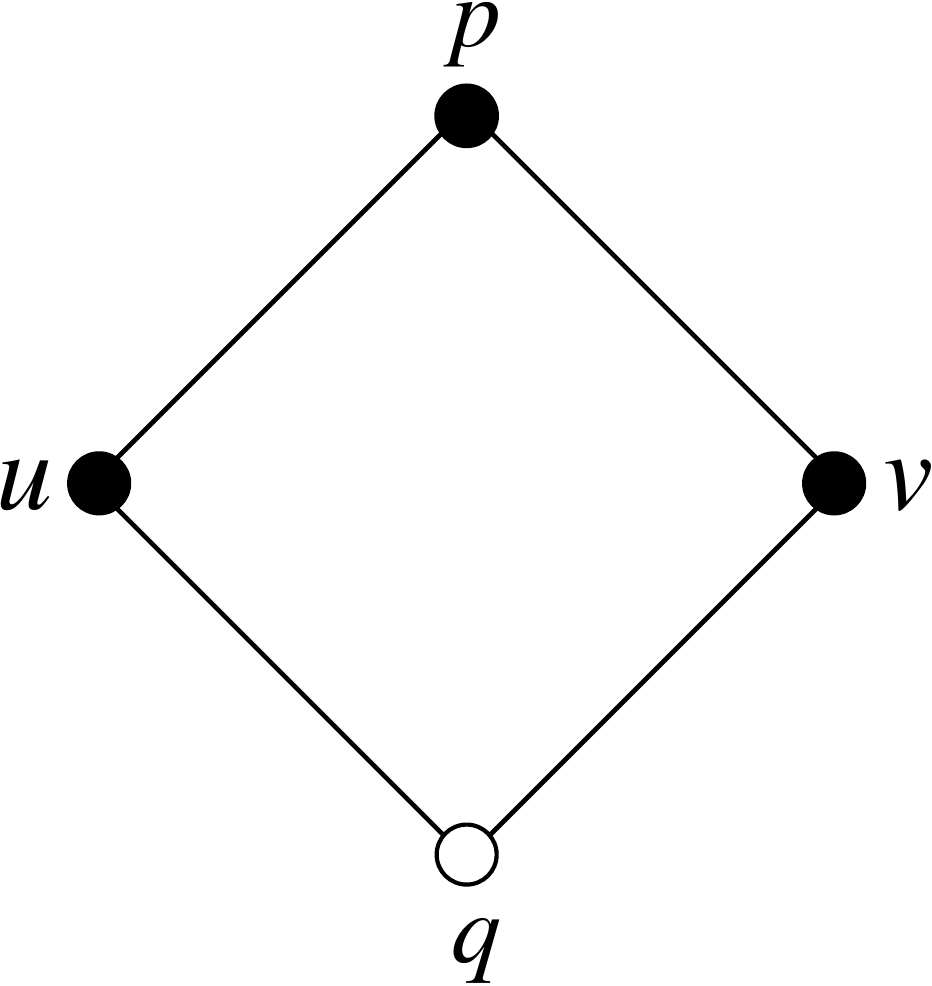}
    \caption{Example of locating an intruder in an IC}
    \label{fig:ic-locate-example}
\end{figure}

As an example, consider the graph presented in Figure~\ref{fig:ic-locate-example}, and the optimal (minimum) identifying code $S$ indicated by shaded vertices.
Suppose we are given $A = \{u,v\}$.
Then $B = \cap_{v \in A} N[v] = \{p,q\}$.
We observe that $p \in N_S[p] \triangle N_S[q]$, so we may eliminate $p \in N[p]$.
Thus, we have only one remaining candidate: $x = q$.
As another example, suppose we have $A = \{u,p\}$, then $B = \{u,p\}$ and $v \in N_S[u] \triangle N_S[p]$, so we eliminate $p \in N[v]$ giving $x = u$.
It is also possible to have $|B| = 1$, which obviates the process of elimination; for example if $A = \{u,p,v\}$, then $B = \{p\}$ implying $x = p$.
An identifying code which always ensures $|B| = 1$ is known as a \emph{self-identifying code} (SIC), as introduced by Junnila and Laihonen \cite{junn20o}.

\begin{definition}[\cite{junn20o}]\label{def:sic}
A set $S \subseteq V(G)$ is an SIC if for any $x \in V(G)$, $N_S[x] \neq \varnothing$ and $\cap_{v \in N_S[x]} N[v] = \{x\}$.
\end{definition}

\begin{theorem}[\cite{junn20o}]\label{theo:sic-char-theirs}
A set $S \subseteq V(G)$ is an SIC if and only if $N_S[u] - N_S[v] \neq \varnothing$ for all distinct $u,v \in V(G)$.
\end{theorem}

In addition to the IC and SIC parameters, other works have also explored fault-tolerant variants of identifying codes.
For instance, the \emph{redundant identifying code} (RED:IC) is an IC which tolerates the removal of any single detector \cite{}.
The \emph{error-detecting identifying code} (DET:IC) is an IC which tolerates one false negative reading from a detector \cite{}.
Similarly, the \emph{error-correcting identifying code} (ERR:IC) is an IC which tolerates any single error (false negative or false positive) from a detector \cite{}.

% As introduced by V. Junnila, and T. Laihonen. \cite{junn20o}  
% ``a certain type of identifying codes, for which the detection of malfunctioning sensors is easy'' and based on these codes they designed a collection of codes tolerant against malfunctions.

% Junnila and Laihonen \cite{junn20o}) characterized SIC: A code is self-identifying if for any distinct $u, v$ we have $N_S[u] \setminus N_S[v] \neq 0$.

% Notation:
% For a dominating set $S \subseteq V(G)$ and $u \in V(G)$, we let $N_S[u] = N[u] \cap S$ denote the \emph{dominators} of $u$ and $dom[u] = |N_S[u]|$. 

For convenience of notation, two distinct vertices $u,v \in V(G)$ are said to be $k$-distinguished by a detector set $S \subseteq V(G)$ if $|N_S[u] - N_S[v]| + |N_S[v] - N_S[u]| \ge k$.
Vertices $u$ and $v$ are said to be $\sharp{k}$-distinguished (or $k$-sharp-distinguished) by $S$ if $\max(|N_S[u] - N_S[v]|, |N_S[v] - N_S[u]|) \ge k$.
Similarly, $u$ and $v$ are said to be $\bisharp{k}$-distinguished (or $k$-bisharp-distinguished) by $S$ if $\min(|N_S[u] - N_S[v]|, |N_S[v] - N[u]|) \ge k$.
In many other works on identifying codes, $k$-distinguishing is often defined as $|N_S[u] \triangle N_S[v]| \ge k$, and the $\sharp{k}$ and $\bisharp{k}$ requirements are typically split into disjunctions and conjunctions, respectively.
In this paper, we have provided alternative definitions to ease comparisons among these distinguishing requirements; for instance, we see that each requirement differs only in a single binary operator.

\begin{figure}[ht]
    \centering
    \begin{tabular}{ccc}
        \includegraphics[width=0.2\textwidth]{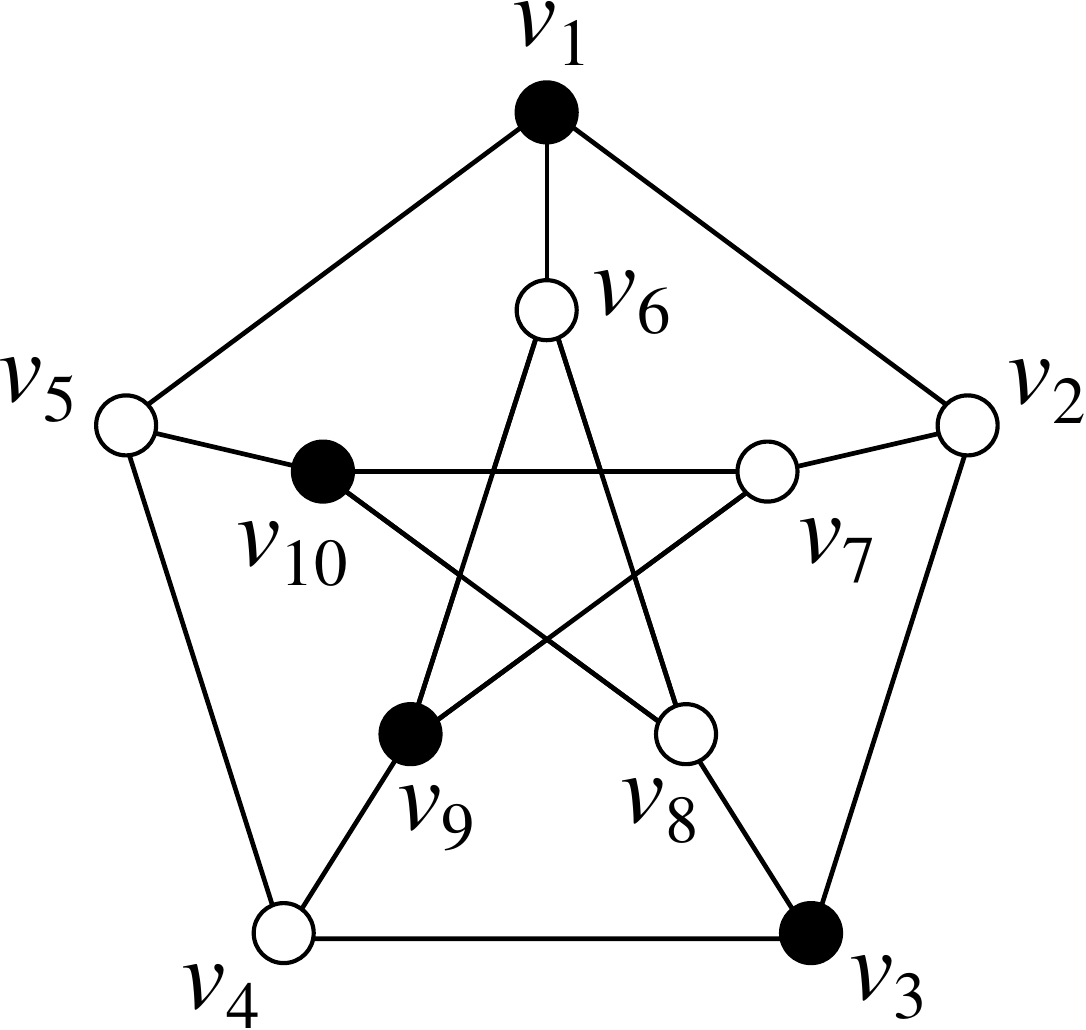} & \includegraphics[width=0.2\textwidth]{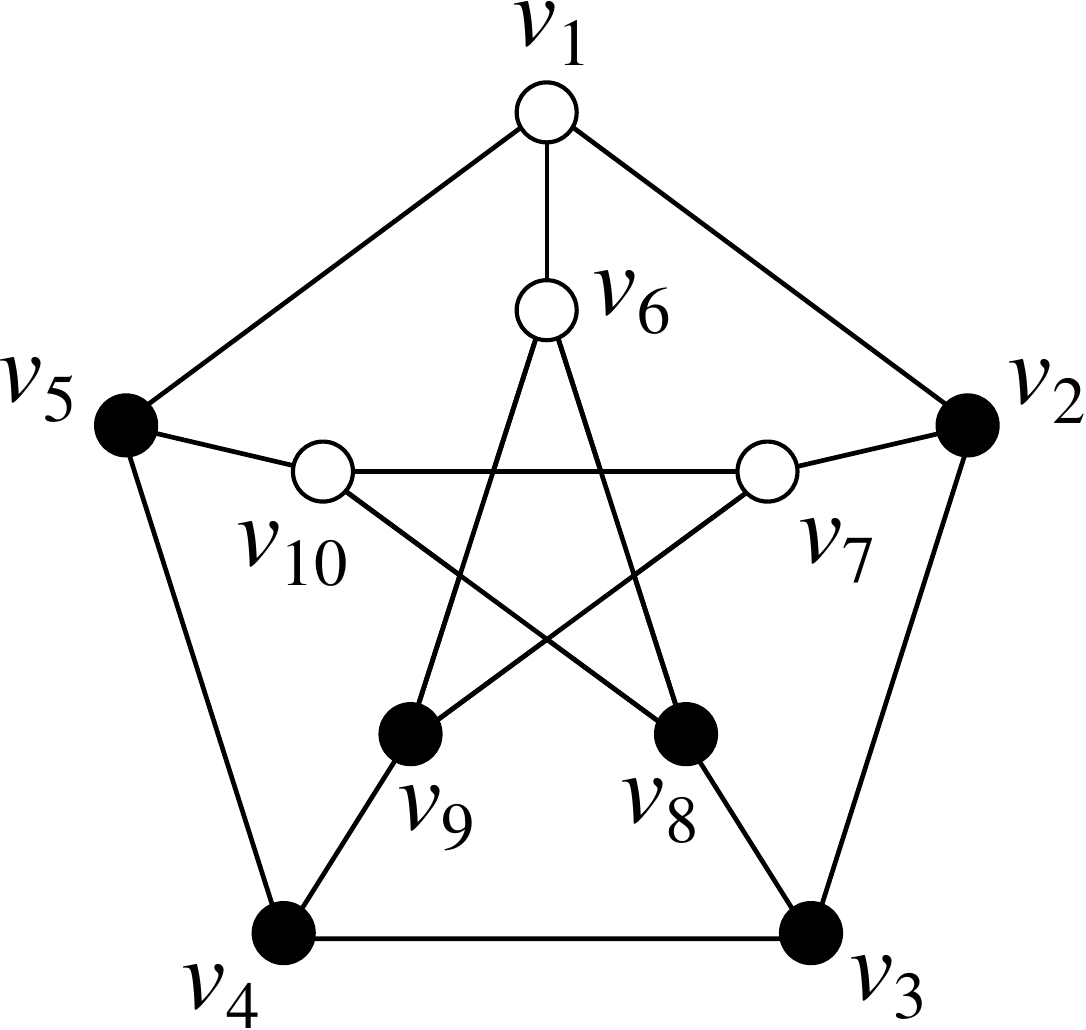} & \includegraphics[width=0.2\textwidth]{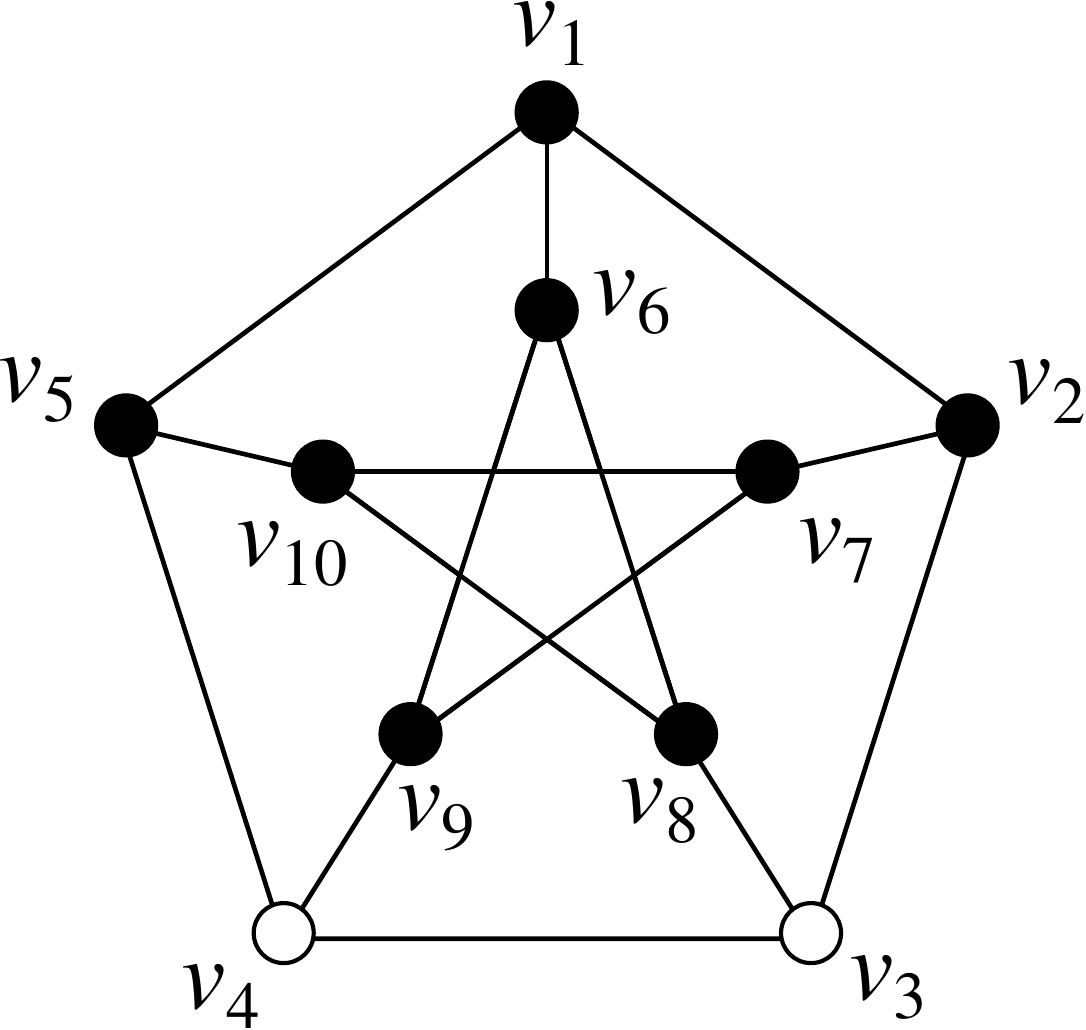} \\ (a) & (b) & (c)
    \end{tabular} \\[1em]
    \begin{tabular}{cc}
        \includegraphics[width=0.2\textwidth]{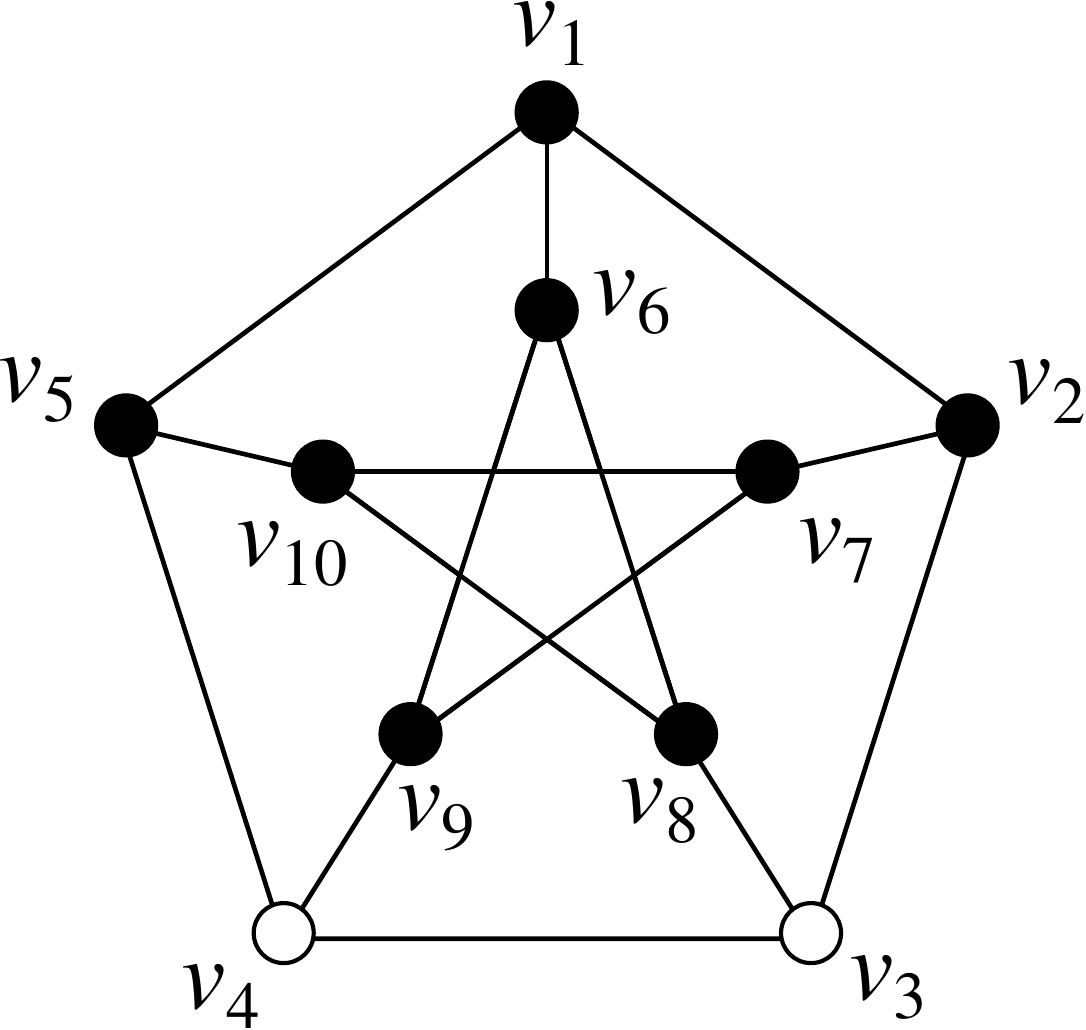} & \includegraphics[width=0.2\textwidth]{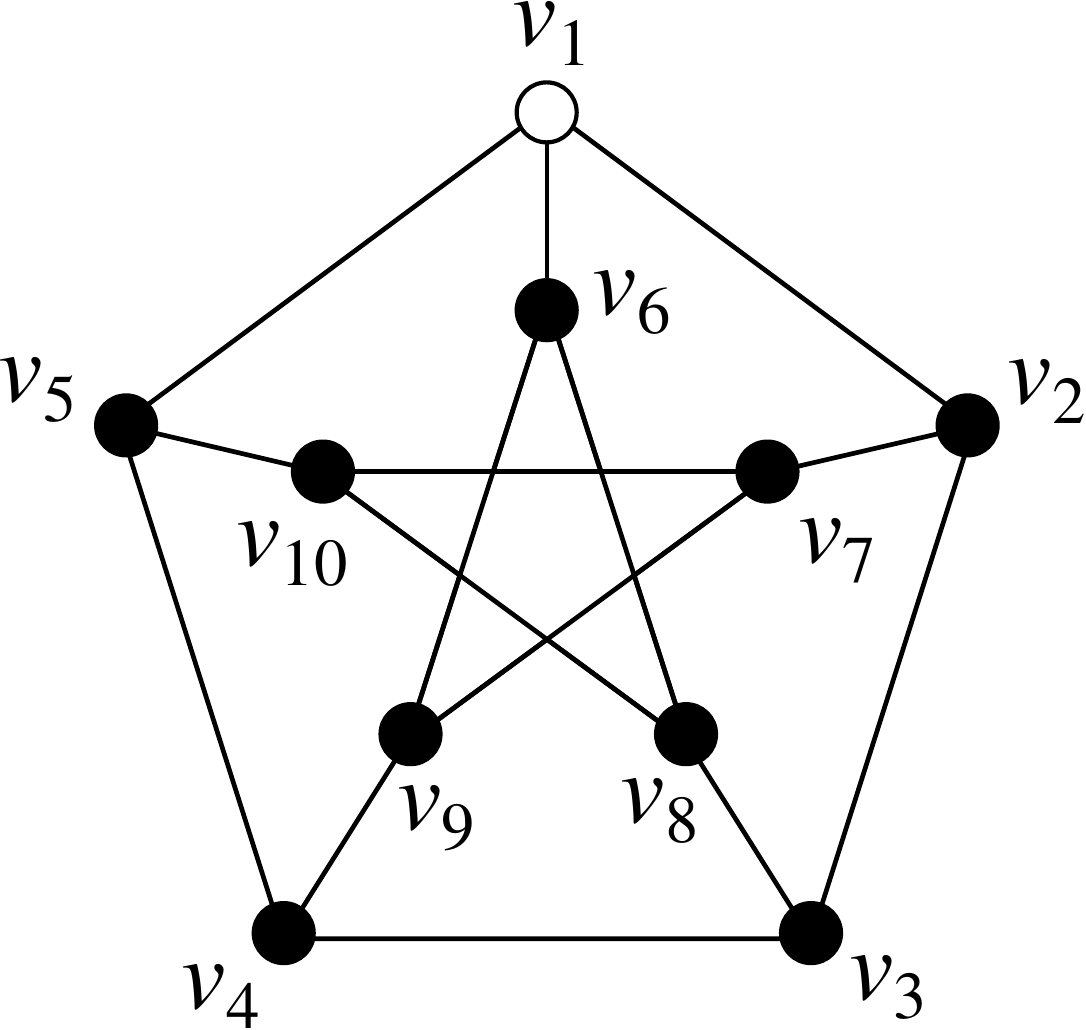} \\ (d) & (e)
    \end{tabular}
    \caption{Examples of optimal IC (a), RED:IC, (b), DET:IC (c), SIC (d), and ERR:IC (e) solutions on the Petersen graph}
    \label{fig:params-on-petersen}
\end{figure}

Table~\ref{tab:ft-ic-cmp} presents characterizations of several identifying code variants in terms of these unified distinguishing notations, and Figure~\ref{fig:params-on-petersen} shows optimal solutions for those parameters on Petersen graph.
The notation $\textrm{SIC}(G)$ denotes the minimum cardinality of an SIC on $G$; for the Petersen graph, we observe that $\textrm{SIC}(G) = 8$.
Similarly, the notation $\textrm{SIC\%}(G)$ denotes the minimum density of an SIC on $G$, or more formally $\limsup_{r \to \infty}\frac{|B_r(c) \cap S|}{|B_r(c)|}$ for any $c \in V(G)$, where $B_r(c)$ denotes the ball of radius $r$ about $c$, $B_r(c) = \{ v \in V(G) : d(c,v) \le r \}$.
This density-based metric is more useful for discussing optimal SICs on infinite graphs, but it can still be used for finite graphs; for instance, we observe that for the Petersen graph $\textrm{SIC\%}(G) = \frac{8}{10}$.

\begin{table}[ht]
    \centering
    { % scope needed for setlength
        \setlength\extrarowheight{0.2em}
        \begin{tabular}{|c|c|c|}
            \hline \textbf{Identifying Code} & \textbf{Domination Requirement} & \textbf{Distinguishing Requirement} \\[0.2em]\hline
            IC \cite{karp98a} & $|N_S[u]| \ge 1$ & $|N_S[u] - N_S[v]| + |N_S[v] - N_S[u]| \ge 1$ \\[0.2em]\hline
            SIC \cite{junn20o} & $|N_S[u]| \ge 1$ & $\min(|N_S[u] - N_S[v]|, |N_S[v] - N_S[u]|) \ge 1$ \\[0.2em]\hline
            RED:IC \cite{jean24a} & $|N_S[u]| \ge 2$ & $|N_S[u] - N_S[v]| + |N_S[v] - N_S[u]| \ge 2$ \\[0.2em]\hline
            DET:IC \cite{jean24b} & $|N_S[u]| \ge 2$ & $\max(|N_S[u] - N_S[v]|, |N_S[v] - N_S[u]|) \ge 2$ \\[0.2em]\hline
            ERR:IC \cite{jean22b} & $|N_S[u]| \ge 3$ & $|N_S[u] - N_S[v]| + |N_S[v] - N_S[u]| \ge 3$ \\[0.2em]\hline  
        \end{tabular}
    }
    \caption{Characterizations of various fault-tolerant identifying codes.}
    \label{tab:ft-ic-cmp}
\end{table}

% \begin{theorem}\label{theo:sic-char}
% A detector set $S \subseteq V(G)$ is an SIC for $G$ if and only if for all $v \in V(G)$, $|N_S[v]| \ge 2$ and for all distinct $u,v \in V(G)$, $|N_S[u] - N_S[v]| \ge 1$ and $|N_S[v] - N_S[u]| \ge 1$.
% \end{theorem}

% Characterization
% Extremal graphs
% Extremal cubic graphs

% Redundant SIC
% RED:SIC(SQR) <= 4/5
% RED:SIC(KNG) <= 2/3
% RED:SIC(HEX) <= 1
% RED:SIC(TRI) <= 3/4

% Error detecting SIC (haven't looked at it yet)
% DET:SIC(SQR) <= 
% DET:SIC(KNG) <= 
% DET:SIC(HEX) <= 
% DET:SIC(TRI) <= 

% Error correcting SIC (haven't looked at it yet)
% ERR:SIC(SQR) <= 
% ERR:SIC(KNG) <= 
% ERR:SIC(HEX) <= 
% ERR:SIC(TRI) <= 

In the next section, we discuss the existence criteria for a self-identifying code. In Section~\ref{sec:npc}, we prove the problem of determining the minimum cardinality of SIC, i.e., $\textrm{SIC}(G)$, for an arbitrary graph $G$ is NP-complete like many existing detection systems \cite{cohe01a, colb87a, jean23a, seo10}.
In Section~\ref{sec:special-classes} we investigate minimum-sized SICs in classes of graphs, including cubic graphs and infinite grids.

\FloatBarrier
\section{Properties and Existence Criteria}\label{sec:existence}

\begin{definition}[\cite{fouc16a}]
Two distinct vertices $u,v \in V(G)$ are said to be \emph{twins} if $N[u] = N[v]$ (\emph{closed-twins}) or $N(u) = N(v)$ (\emph{open-twins}).
\end{definition}

\begin{definition}
Two distinct vertices $u,v \in V(G)$ are said to be \emph{semi-twins} if $N[u] \sim N[v]$ (\emph{semi-closed-twins}) or $N(u) \sim N(v)$ (\emph{semi-open-twins}), where $A \sim B$ denotes $A \subseteq B$ or $B \subseteq A$.
\end{definition}

It is easy to see $G$ has an IC if and only if $G$ has no closed-twins.
Similarly, it is easy to see that $G$ has an SIC if and only if $G$ has no semi-closed-twins.
As a particular example, if $G$ has a vertex $v$ with $deg(v) = 1$ (i.e., a leaf) and $n \ge 2$, then $G$ has no SIC.
On the opposite extreme, if $G$ has a vertex $v$ with $N[v] = V(G)$ (or equivalently $deg(v) = n - 1$) and $n \ge 2$, then $G$ has no SIC.

\begin{observation}
A graph $G$ has an SIC if and only if $G$ has no semi-closed-twins.
\end{observation}

\begin{corollary}
Any graph $G$ with $n \ge 2$ that permits SIC must have $\delta(G) \ge 2$.
\end{corollary}

It can be shown exhaustively that every graph on $n \le 3$ vertices contains a semi-closed-twin.
The smallest semi-closed-twin-free graph is $C_4$, which is thus the smallest graph permitting SIC.

\begin{corollary}
The smallest graph permitting SIC is $C_4$.
\end{corollary}

Clearly, as all twins are semi-twins, we find that the existence of SIC implies the existence of IC.
Moreover, from Table~\ref{tab:ft-ic-cmp}, we find that any SIC is also an IC.
Although the characterization of SIC presented in Table~\ref{tab:ft-ic-cmp} requires only 1-domination, for any graph with $n \ge 2$ the $\bisharp{1}$-distinguishing requirement necessitates 2-domination.

\begin{theorem}[\cite{jean24b}]\label{theo:delta-g-sic-redic}
If $G$ has $\delta(G) \ge 1$ and $S \subseteq V(G)$ is 1-dominating and $\bisharp{1}$-distinguishing, then $S$ is 2-dominating.
\end{theorem}
\begin{proof}
Suppose there is some $v \in V(G)$ which is only 1-dominated.
If $v \notin S$, then the fact that $S$ is 1-dominating implies there is some $u \in N_S(v)$; or if $v \in S$, then the fact that $\delta(G) \ge 1$ implies there is some $u \in N(v) \setminus S$.
In either case we see that $u$ and $v$ cannot be 1-bisharp-distinguished, a contradiction, completing the proof.
\end{proof}

\begin{corollary}
If $G$ has $n \ge 2$, then any SIC is also a RED:IC.
\end{corollary}

\begin{theorem}
If $G$ is $k$-regular, then SIC exists iff IC exists.
\end{theorem}
\begin{proof}
Clearly, if SIC exists, then IC exists.
For the converse, we are given that $G$ permits an IC; this implies that $S = V(G)$ is an IC, and we will show $S$ is also an SIC.
Because $S$ is an IC, any $v \in V(G)$ must have $|N_S[v]| \ge 1$, which satisfies the domination requirement of SIC as per Table~\ref{tab:ft-ic-cmp}.
Next, because $S$ is an IC, any distinct $u,v \in V(G)$ must have $|N_S[u] \triangle N_S[v]| \ge 1$, which implies the existence of some $p \in (N_S[u] - N_S[v]) \cup (N_S[v] - N_S[u])$.
Without loss of generality, we will assume $p \in N_S[u] - N_S[v]$.
Because $G$ is regular, the existence of $p$ implies the existence of some $q \in N_S[v] - N_S[u]$.
Thus, $u$ and $v$ are $\bisharp{1}$-distinguished, completing the proof.
\end{proof}

\FloatBarrier
\section{NP-completeness of Self Identifying Codes}\label{sec:npc}

It has been shown that many graphical parameters related to detection systems, such as finding the cardinality of the smallest IC, RED:IC, DET:IC, and ERR:IC are NP-complete problems \cite{char03b, jean22b,jean24a, jean24b}.
We will now prove that the problem of determining the smallest SIC for an arbitrary graph is also NP-complete.
For additional information about NP-completeness, see Garey and Johnson \cite{np-complete-bible}.

\npcompleteproblem{3-SAT}{Let $X$ be a set of $N$ variables.
Let $\psi$ be a conjunction of $M$ clauses, where each clause is a disjunction of three literals from distinct variables of $X$.}{Is there is an assignment of values to $X$ such that $\psi$ is true?}

\npcompleteproblem{Self Identifying Code (SIC)}{A graph $G$ and integer $K$ with $4 \le K \le |V(G)|$.}{Is there an SIC set $S$ with $|S| \le K$? Or equivalently, is $\textrm{SIC}(G) \le K$?}

\begin{lemma}\label{lem:p2deg2}
If $S$ is an SIC for $G$ and $u,v \in V(G)$ with $u \in N(v)$ and $\textrm{deg}(u) = \textrm{deg}(v) = 2$, then $N[u] \cup N[v] \subseteq S$.
\end{lemma}
\begin{proof}
We know that $deg(u) = deg(v) = 2$; let $p \in N(u) - \{v\}$ and $q \in N(v) - \{u\}$.
In order to distinguish $u$ and $v$, it must be that $p \neq q$ and $\{p,q\} \subseteq S$.
Next, we see that in order to distinguish $u$ and $p$ we require $v \in S$; by symmetry, $u \in S$ as well, completing the proof.
\end{proof}

\begin{figure}[ht]
    \centering
    \includegraphics[width=0.4\textwidth]{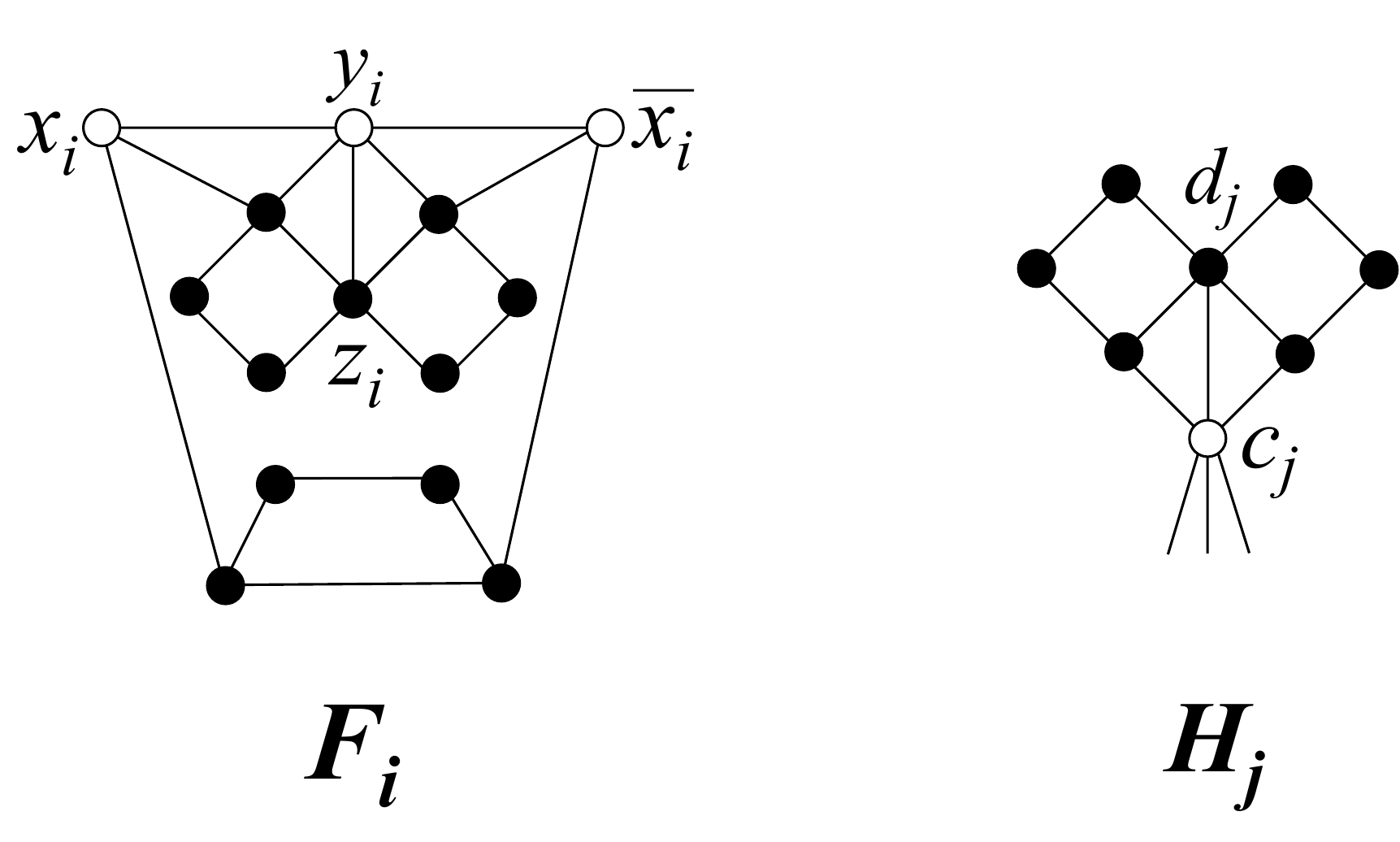}
    \caption{Variable and clause graphs}
    \label{fig:variable-clause}
\end{figure}

\begin{figure}[ht]
    \centering
    \includegraphics[width=0.95\textwidth]{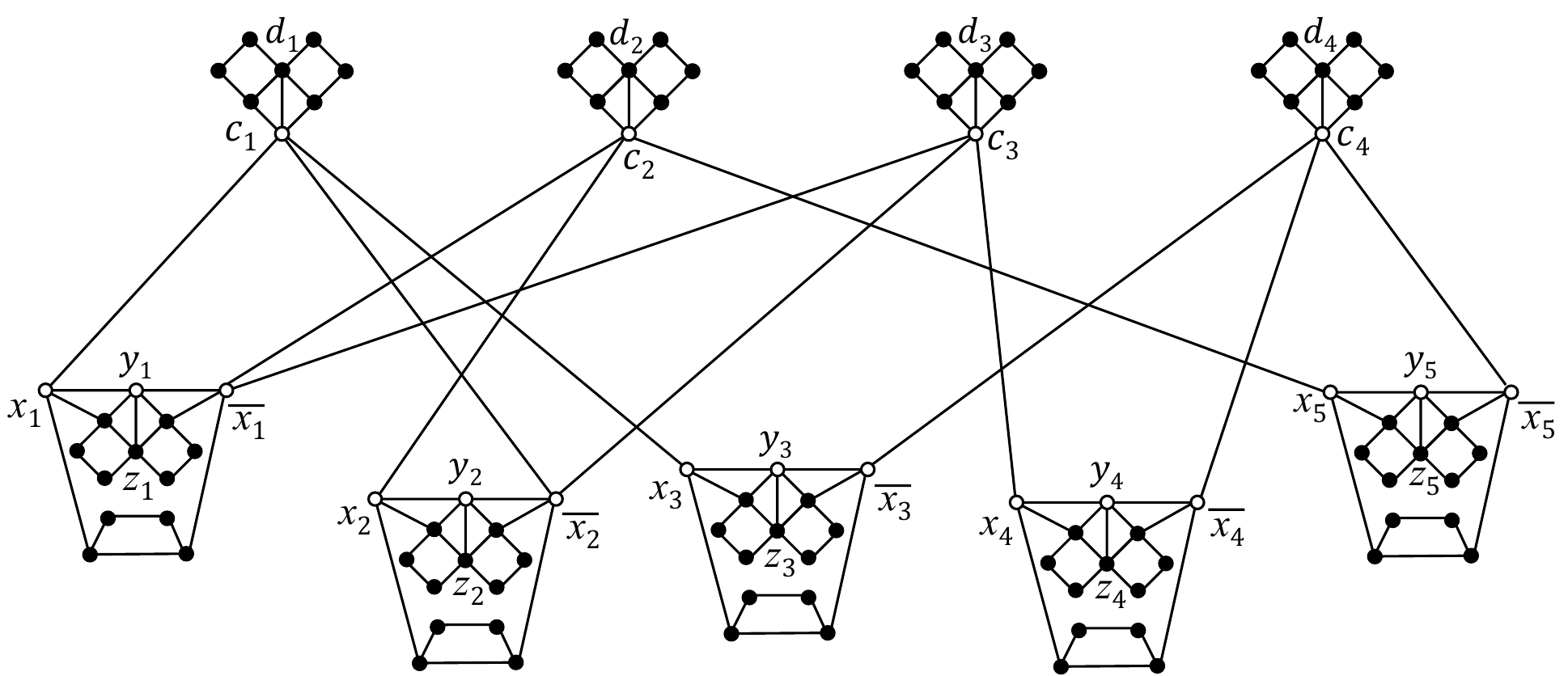}
    \caption{\begin{tabular}[t]{l} Construction of $G$ from $(x_1 \lor \overline{x_2} \lor x_3) \land (\overline{x_1} \lor x_2 \lor x_5) \land (\overline{x_1} \lor \overline{x_2} \lor x_4) \land (\overline{x_3} \lor \overline{x_4} \lor \overline{x_5})$ \protect\\ with $N = 4$, $M = 3$, $K = 109$  \end{tabular}}
    \label{fig:example-clause}
\end{figure}

\begin{theorem}
The SIC problem is NP-complete.
\end{theorem}
\begin{proof}
Clearly, SIC is NP, as every possible candidate solution can be generated nondeterministically in polynomial time (specifically, $O(n)$ time), and each candidate can be verified in polynomial time using Theorem~\ref{theo:sic-char-theirs}.
To complete the proof, we will now show a reduction from 3-SAT to SIC.

Let $\psi$ be an instance of the 3-SAT problem with $M$ clauses on $N$ variables.
We will construct a graph, $G$, as follows.
For each variable $x_i$, create an instance of the $F_i$ graph shown in Figure~\ref{fig:variable-clause}; this includes a vertex for $x_i$ and its negation $\overline{x_i}$.
For each clause $c_j$ of $\psi$, create an instance of the $H_j$ graph shown in Figure~\ref{fig:variable-clause}.
For each clause $c_j = \alpha_j \lor \beta_j \lor \gamma_j$, create an edge from the $y_j$ vertex to $\alpha_j$, $\beta_j$, and $\gamma_j$ from the variable graphs, each of which is either some $x_i$ or $\overline{x_i}$; for an example, see Figure~\ref{fig:example-clause}.
The resulting graph has precisely $14N + 8M$ vertices and $21N + 14M$ edges, and can be constructed in polynomial time.
To complete the problem instance, we define $K = 12N + 7M$.

Suppose $S \subseteq V(G)$ is an SIC on $G$ with $|S| \le K$.
By Lemma~\ref{lem:p2deg2}, we require at least all of the $11N + 7M$ detectors which are shaded in Figure~\ref{fig:variable-clause}.
Additionally, in order to distinguish each $y_i$ from $z_i$, we require either $x_i \in S$ or $\overline{x_i} \in S$.
Therefore, $|S| \ge 12N + 7M$; if $|S| = 12N + 7M$, then each clause vertex $c_j$ must be connected to at least one detector variable vertex (some $x_i$ or $\overline{x_i}$) under a proper truth value assignment for each variable, so we have a solution to 3-SAT problem $\psi$.

For the converse, suppose we have a solution to the 3-SAT problem $\psi$; we will show that there is an SIC, $S$, on $G$ with $|S| \le K$.
We construct $S$ by first including all of the $11N + 7M$ vertices as shown in Figure~\ref{fig:variable-clause}.
Next, for each variable, $x_i$, if $x_i$ is true in our solution to $\psi$, then we let the vertex $x_i \in S$; otherwise, we let $\overline{x_i} \in S$.
Thus, the fully-constructed $S$ has $|S| = 12N + 7M = K$, and it can be shown that $S$ satisfies Theorem~\ref{theo:sic-char-theirs}, so $S$ is an SIC for $G$ with $|S| \le K$, completing the proof.
\end{proof}

\FloatBarrier
\section{SIC for Special Classes of Graphs}\label{sec:special-classes}

\subsection{Hypercubes}

\begin{figure}[ht]
    \centering
    \includegraphics[width=0.5\textwidth]{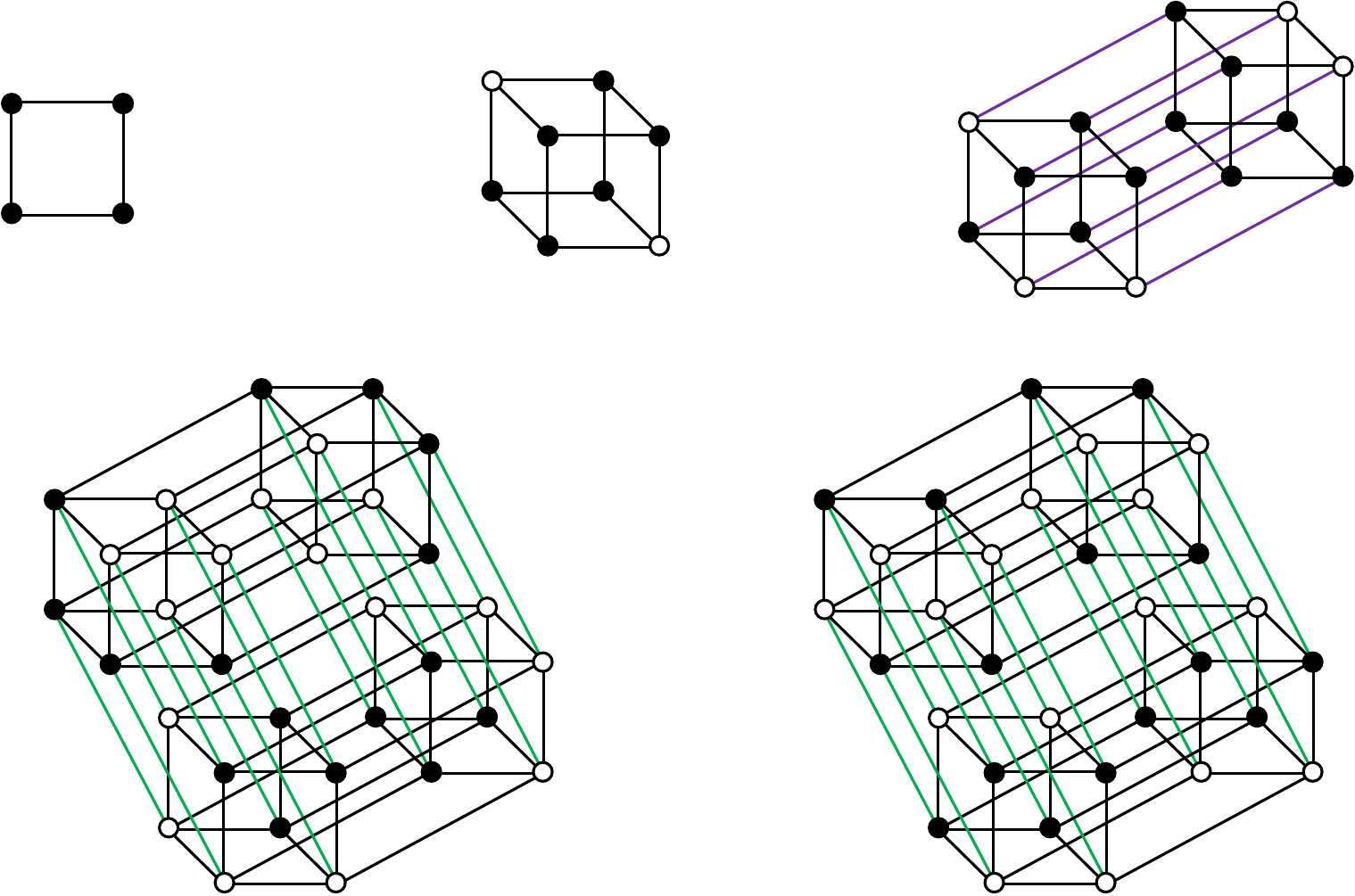}
    \caption{Optimal SIC for $Q_n$ with $n \le 5$}
    \label{fig:hypercubes}
\end{figure}

Let $Q_n = P_2^n$, where $G^n$ denotes repeated application of the $\square$ operator, be the hypercube in $n$ dimensions.
If $S$ is an SIC on $Q_n$ for $n \ge 2$, then we can duplicate the vertices to produce a new SIC of size $2|S|$ on $Q_{n+1} = Q_n \square P_2$; thus, $\textrm{SIC\%}(Q_n)$ is a non-increasing sequence in terms of $n$.
We have found that $\textrm{SIC\%}(Q_5) = \frac{1}{2}$, which serves as an upper bound for the minimum density of SICs in larger hypercubes.
Figure~\ref{fig:hypercubes} shows an optimal SIC for each of the hypercubes on $n \le 5$ dimensions.
From this, we observe that $\textrm{SIC\%}(Q_2) = \frac{4}{4}$, $\textrm{SIC\%}(Q_3) = \frac{6}{8}$, $\textrm{SIC\%}(Q_4) = \frac{11}{16}$, and $\textrm{SIC\%}(Q_5) = \frac{16}{32}$.
Moreover, the presented solutions on $Q_2$, $Q_3$, and $Q_4$ are unique up to isomorphism, and the two distinct solutions for $Q_5$ are the only two minimum solutions up to isomorphism.

% finite ladders

% infinite ladder: 2/3

% finite cylinders

% tori
% $C3 \square C3$

% \subsection{Finite tori}

% \begin{figure}[ht]
%     \centering
%     \includegraphics[width=0.9\textwidth]{temp-fig/Screenshot from 2023-09-20 18-03-23.png}
%     \caption{Caption}
%     \label{fig:enter-label}
% \end{figure}

% \FloatBarrier

% {\color{red} EDIT}

% \FloatBarrier
% \subsection{Uniqueness of minimum solutions in $C_i \square C_j$}

% \begin{conjecture}
% The torus $C_i \square C_j$ (where $i, j \ge 3$) has a unique minimum SIC if and only if $\{ (i, j), (j, i) \} \subseteq \{(3,3), (3,5)\} \cup \{(4, x) : x = 2t, t \in \mathbb{N} \} \cup \{(x, x) : x = 2t, t \in \mathbb{N}\} \cup \{ (x, y) : x + y = 2t + 1, t \in \mathbb{N}, x \ge 5, y \ge 5 \}$.
% \end{conjecture}

% \FloatBarrier

\subsection{SIC in cubic graphs}\label{sec:sic-cubic}

In this section, we consider cubic graphs which permit the SIC parameter.
In Section~\ref{sec:sic-cubic-upper}, we present the upper bound for the value of $\textrm{SIC}(G)$ in cubic graphs, characterize the graphs achieving this extremal value, and provide an infinite family of cubic graphs achieving the bound.
In section~\ref{sec:sic-cubic-lower}, we present the lower bound for the value of $\textrm{SIC}(G)$ in cubic graphs and an infinite family of cubic graphs achieving this bound.

\subsubsection{Upper Bound on SIC(G)}\label{sec:sic-cubic-upper}

\begin{theorem}\label{theo:sic-cubic-x-y}
Let $G$ be a cubic graph permitting SIC with $n \ge 8$. If $x \in V(G)$ is not adjacent to a triangle (not containing itself), then there exists $y \in B_2(x)$ where $y$ is both not adjacent to a triangle and is not a twin.
\end{theorem}
\begin{proof}
Let $N(x) = \{a,b,c\}$.
If $x$ is not a twin, then we could let $y = x$ and be done; otherwise, we will assume that $x$ must be a twin.
Because $G$ permits SIC by assumption, $G$ cannot contain closed-twins; thus, we know that there must be some $x' \in V(G) - \{x\}$ with $N(x') = N(x)$.
We see that $ab \notin E(G)$, as otherwise we would create closed-twins; by symmetry $ac,bc \notin E(G)$ as well.
Next, we see that if $a$ is a twin, then it must be a twin with $b$ and/or $c$ due to sharing $x$ and $x'$.
The only way to have $a$, $b$, and $c$ all be twins is to have $G = K_{3,3}$, which would contradict the assumption that $n \ge 8$.
Thus, there must be at least one vertex in $N(x)$ which is not a twin; without loss of generality let this be vertex $a$.
Because $a$ is not a twin, there must be some $p \in N(a) - \{x,x'\}$ with $p \notin N(b) \cup N(c)$.
Let $N(p) = \{a,u,v\}$.
If $uv \notin E(G)$, then we could let $y = a$ and be done; otherwise we will assume that $uv \in E(G)$.
Finally, we see that $u$ and $v$ cannot be part of a triangle which excludes $p$, so we can let $y = p$, completing the proof.
\end{proof}

\begin{theorem}\label{theo:sic-cubic-tri-n-1}
Let $G$ be a cubic graph permitting SIC with $n \ge 8$. If $x \in V(G)$ is not adjacent to a triangle (not containing itself), then there exists $y \in B_2(x)$ such that $S = V(G) - \{y\}$ is an SIC for $G$.
\end{theorem}
\begin{proof}
By Theorem~\ref{theo:sic-cubic-x-y}, let $y \in B_2(x)$ where $y$ is both not adjacent to a triangle (not containing itself) and is also not a twin.
Let $N(y) = \{a,b,c\}$.
To demonstrate that $S = V(G) - \{y\}$ is an SIC for $G$, we will show that every vertex is at least 1-dominated and all pairs are 1-bisharp distinguished.
Because $G$ is cubic and $y$ is the only non-detector, it is trivially seen that every vertex is at least 3-dominated.
Thus, we need only show that an arbitrary vertex pair, $u,v \in V(G)$, is distinguished.
Firstly, if $u,v \notin N[y]$, then the assumption that $G$ permits an SIC implies that $u$ and $v$ are still distinguished regardless of $y$ being a non-detector.
Next, suppose $u \in N[y]$ but $v \notin N[y]$.
If $u = y$, then $u$ and $v$ must be 1-bisharp distinguished because we know $y$ is not a twin; otherwise, by symmetry we can assume that $u = a$.
If $u \notin N(v)$, then $u$ and $v$ would be 1-bisharp distinguished from their own vertices, respectively; otherwise, we can assume that $u \in N(v)$.
From here, we see that $N(u) \cap N(v) = \varnothing$, as otherwise $y$ would be adjacent to a triangle, a contradiction; thus, we see that $u$ and $v$ are 1-bisharp distinguished.
Finally, suppose that $u,v \in N[y]$.
If $u \in N(v)$, then the fact that SIC forbids closed-twins and $y$ is the only non-detector implies that $u$ and $v$ must still be distinguished; otherwise, we can assume that $u \notin N(v)$.
Because $u \notin N(v)$, we know that $\{u,v\} \cap \{y\} = \varnothing$, implying $u,v \in S$ and that $(u,v)$ must be at least 1-bisharp distinguished, completing the proof.
\end{proof}

\begin{theorem}\label{theo:sic-cubic-everything-tri}
Let $G$ be a cubic graph permitting SIC with $n \ge 8$. Then $\textrm{SIC}(G) = n$ if and only if every vertex is adjacent to a triangle (not containing itself).
\end{theorem}
\begin{proof}
Suppose $S$ is an SIC on $G$ and $v \in V(G)$ is adjacent to a triangle $abc$ with $v \in N(a)$ but $v \notin \{a,b,c\}$.
In order to distinguish $(a,b)$, we require $v \in S$.
Therefore, if every vertex in $G$ is adjacent to a triangle (not containing itself), then $\textrm{SIC}(G) = n$.
The inverse is given by Theorem~\ref{theo:sic-cubic-tri-n-1}, completing the proof.
\end{proof}

\begin{theorem}\label{theo:sic-cubic-disjoint-tri}
If $G$ is a cubic graph permitting SIC, then a vertex may be part of at most one triangle.
\end{theorem}
\begin{proof}
Suppose $v$ is part of two distinct triangles, $vab$ and $vxy$.
Because $G$ is cubic, without loss of generality we can assume $a = x$.
Then we find that $v$ and $a$ are closed twins, contradicting the existence of SIC, completing the proof.
\end{proof}

\begin{theorem}\label{theo:cubic-adj-tri-in-tri}
Let $G$ be a cubic graph permitting SIC. Then every vertex is adjacent to a triangle (not containing itself) if and only if every vertex is part of a triangle.
\end{theorem}
\begin{proof}
For the forward direction, suppose $v \in V(G)$ is adjacent to a triangle $abc$ with $v \in N(a)$ but $v \notin \{a,b,c\}$.
We observe that $b$ or $c$ cannot be part of a triangle other than $abc$ without becoming a closed-twin, so the triangle $a$ is adjacent to must contain vertex $v$, completing the forward implication.
For the converse, if every vertex is part of a triangle, then every vertex must also be adjacent to a triangle.
By Theorem~\ref{theo:sic-cubic-disjoint-tri}, a vertex can be part of at most one triangle, implying that every vertex is moreover adjacent to a triangle not containing itself, completing the proof.
\end{proof}

From Theorems \ref{theo:sic-cubic-everything-tri}, \ref{theo:cubic-adj-tri-in-tri}, and \ref{theo:sic-cubic-disjoint-tri}, we find that a cubic graph permitting SIC with $n \ge 8$ has $\textrm{SIC}(G) = n$ if and only if it can be partitioned into a set of disjoint triangles.
This serves as a full characterization of the extremal cubic graphs which require all vertices be detectors.

\begin{corollary}\label{cor:sic-cubic-n-char}
Let $G$ be a cubic graph permitting SIC with $n \ge 8$. Then $\textrm{SIC}(G) = n$ if and only if $G$ can be partitioned into a set of disjoint triangles.
\end{corollary}

\begin{figure}[ht]
    \centering
    \includegraphics[width=0.4\textwidth]{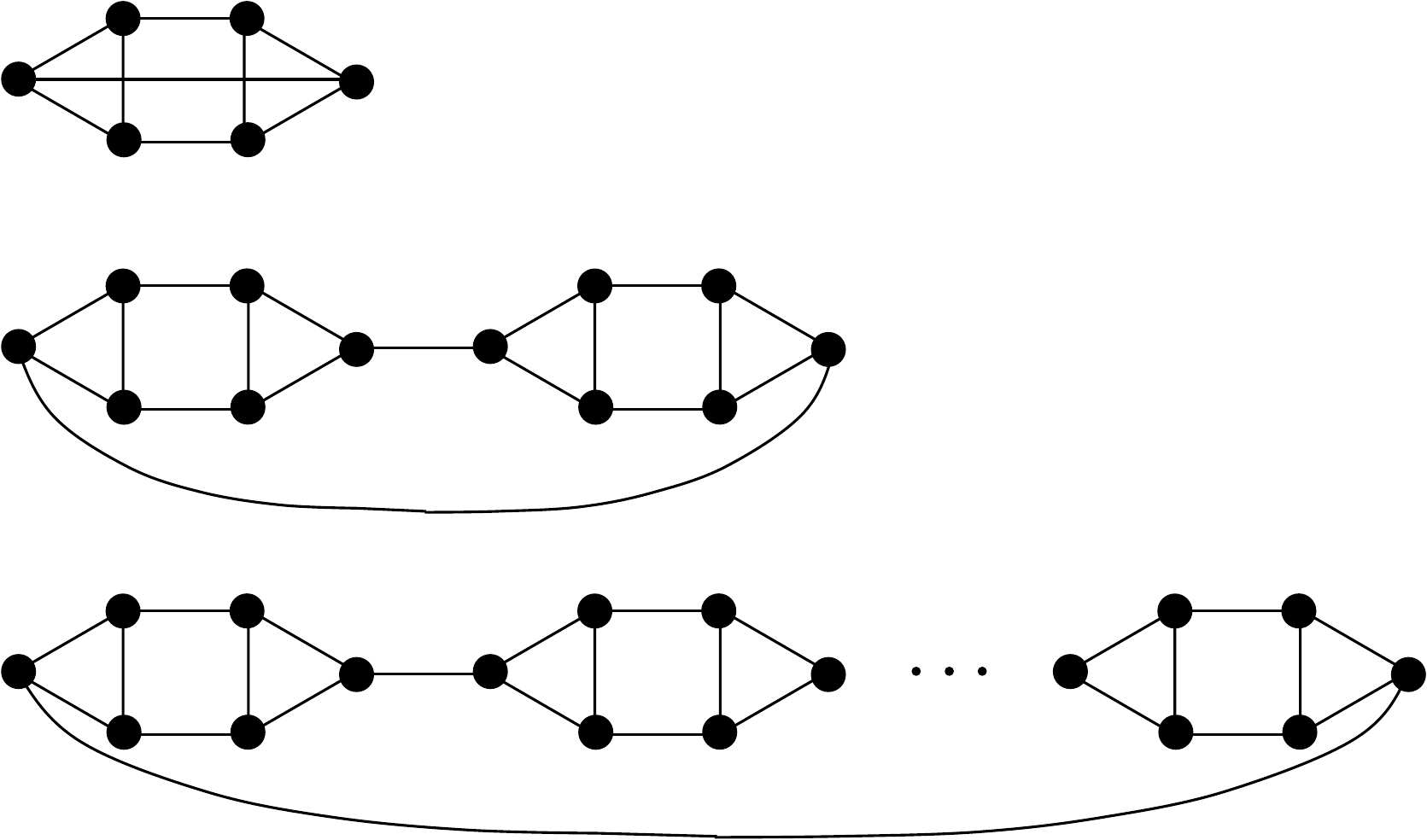}
    \caption{An infinite family of cubic graphs on $n=6k$ with $\textrm{SIC}(G_n) = n$.}
    \label{fig:inf-fam-6n-sic}
\end{figure}

Corollary~\ref{cor:sic-cubic-n-char} provides a full characterization of the cubic graphs with $\textrm{SIC}(G) = n$.
Figure~\ref{fig:inf-fam-6n-sic} presents a particular infinite family of cubic graphs on $n = 6k$ vertices which follows this extremal construction.

\begin{theorem}\label{theo:sic-cubic-n-6k}
Let $G$ be a cubic graph permitting SIC with $n \ge 8$. Then $\textrm{SIC}(G) = n$ implies $n = 6k$.
\end{theorem}
\begin{proof}
We know that $\textrm{SIC}(G) = n$.
By Theorem~\ref{theo:sic-cubic-everything-tri}, this is true if and only if every vertex is adjacent to a triangle (not containing itself).
By Theorem~\ref{theo:cubic-adj-tri-in-tri}, this is true if and only if every vertex is part of a triangle.
By Theorem~\ref{theo:sic-cubic-disjoint-tri}, a vertex can be part of at most one triangle; thus, every vertex must be part of exactly one triangle.
We can then partition $G$ into a collection of $\frac{n}{3}$ disjoint triangles, implying $n$ is divisible by 3.
Because $G$ is cubic, we also know that $n$ must be even.
Therefore, we find that $n = 6k$ for some $k \in \mathbb{N}$, completing the proof.
\end{proof}

\FloatBarrier
\subsubsection{Lower Bound on SIC(G)}\label{sec:sic-cubic-lower}

\begin{theorem}\label{theo:sic-reg-lower-2-k}
If $G$ is a $k$-regular graph with SIC, then $\textrm{SIC}\%(G) \ge \frac{2}{k}$.
\end{theorem}
\begin{proof}
Let $x \in S$ be an arbitrary detector.
We know that $|N[x] \cap S| \ge 3$ in order for $x$ to be distinguished from each vertex in $N(x) \cap S$.
Thus, every detector vertex $x \in S$ must be at least 3-dominated.
Because $|N(x) \cap S| \ge 2$, this implies that $x$ is adjacent to at least two vertices which are at least 3-dominated.
The other $k-2$ detectors could potentially all be non-detectors, in which case they may be only 2-dominated.
In any case, $sh(x) \le \frac{3}{3} + \frac{k-2}{2} = \frac{k}{2}$, completing the proof.
\end{proof}

\begin{figure}[ht]
    \centering
    \includegraphics[width=0.4\textwidth]{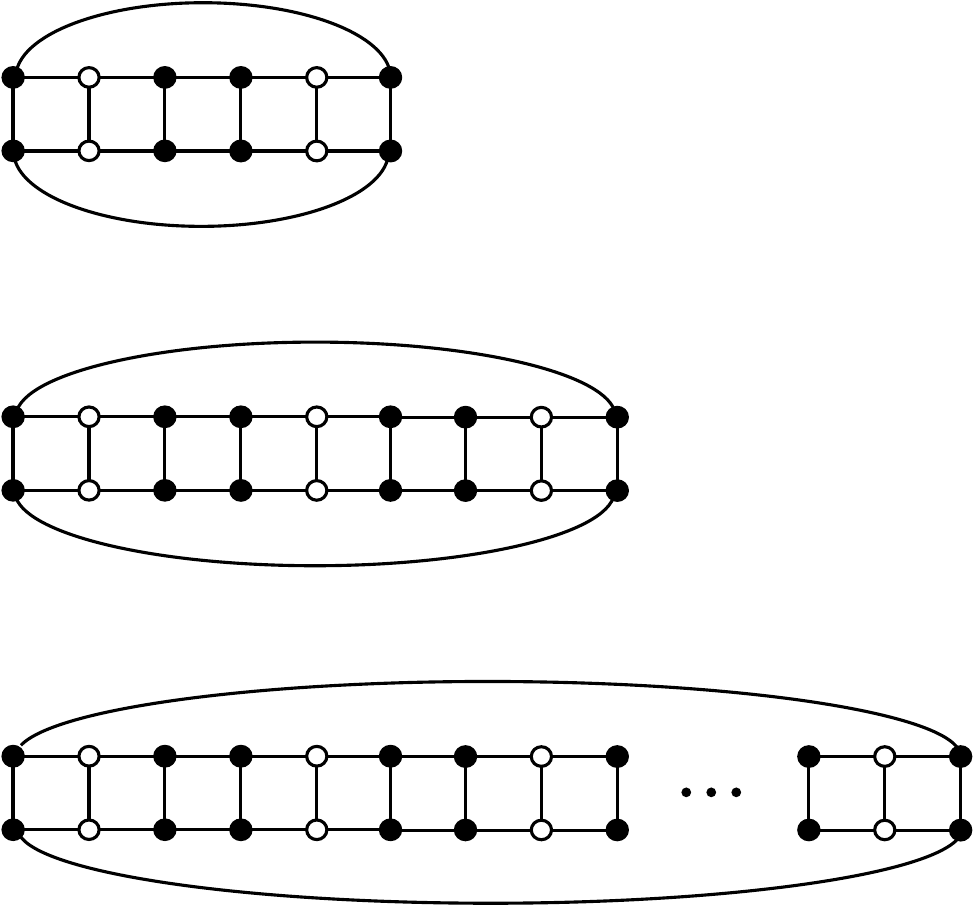}
    \caption{An infinite family of cubic graphs on $n=6k$ for $k \ge 2$ with $\textrm{SIC\%}(G_n) = \frac{2}{3}$.}
    \label{fig:sic-cubic-lower-family}
\end{figure}

From Theorem~\ref{theo:sic-reg-lower-2-k}, we know $\textrm{SIC}(G) \ge \frac{2}{3}$ for a cubic graph $G$, and Figure~\ref{fig:sic-cubic-lower-family} shows an infinite family of cubic graph achieving the bound.

\begin{figure}[ht]
    \centering
    \includegraphics[width=0.45\textwidth]{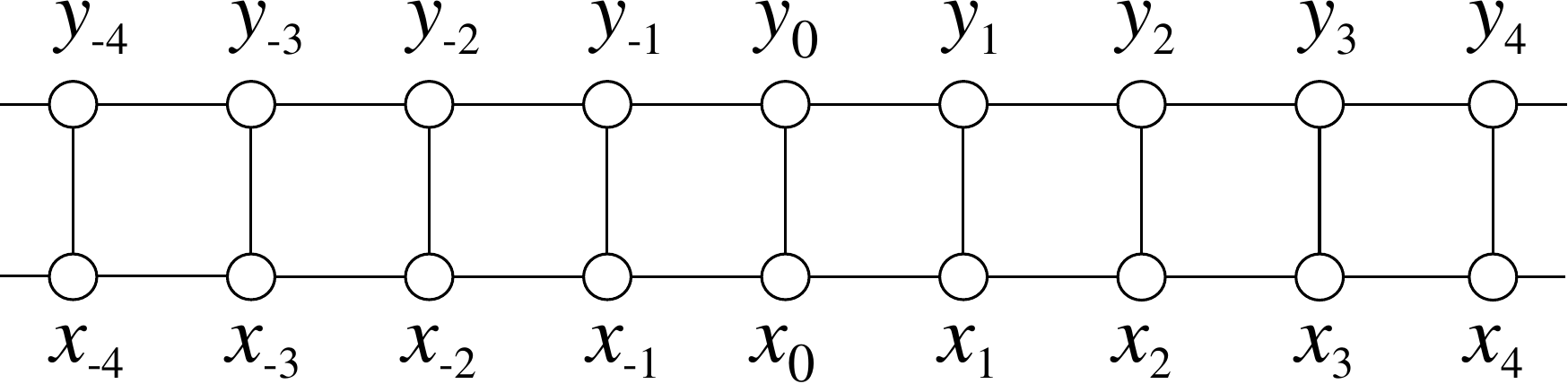}
    \caption{Labeling of vertices in $G = C_{3k} \square P_2$}
    \label{fig:sic-fam-lower-label}
\end{figure}

\begin{observation}\label{obs:sic-no-leaves}
If $S$ is an SIC for $G$ and $v \in S$, then $v$ is at least 3-dominated.
\end{observation}
\begin{proof}
Suppose $N[v] \cap S = \{v,u\}$; then $u$ and $v$ are not distinguished, a contradiction.
\end{proof}

\begin{theorem}\label{theo:sic-fam-lower}
The graph $G = C_{3k} \square P_2$ for $k \ge 2$ has $\textrm{SIC}\%(G) = \frac{2}{3}$ and has a unique solution up to isomorphism as shown in Figure~\ref{fig:sic-cubic-lower-family}.
\end{theorem}
\begin{proof}
Let $G$ have the vertex labeling shown in Figure~\ref{fig:sic-fam-lower-label}.
We will use an association argument with association region $A(x_i) = A(y_i) = N[x_i] \cup N[y_i]$.
To begin, let $x_0 \notin S$ be an arbitrary non-detector.
We see that $x_1 \in S$ is required to distinguish $(x_1, y_2)$, and by symmetry $x_{-1} \in S$ as well.
By applying Observation~\ref{obs:sic-no-leaves} to $x_1$, we require $\{x_2, y_1\} \subseteq S$, and by symmetry $\{x_{-2}, y_{-1}\} \subseteq S$ as well.
Finally, we see that $y_2 \in S$ is required to distinguish $(x_0, y_1)$, and by symmetry $y_{-2} \in S$ as well.
If $y_0 \notin S$, we see that $A(x_0) = A(y_0)$ has a density of $\frac{2}{3}$ and cannot overlap with any other $A(v)$ for any $v \in V(G) - A(x_0)$.
Otherwise, $y \in S$, and we see that we have the same result but with a local density of $\frac{5}{6}$.
Thus, the minimum global density of $S$ in $G$ is $\frac{2}{3}$.
For the uniqueness argument, it suffices to see that the density of $\frac{2}{3}$ can only be achieved by densely covering $V(G)$ with $A(x_{3j}) = A(y_{3j})$ for all $j \in \mathbb{Z}$ (with wrapping around $C_{3k}$), completing the proof.
\end{proof}

Table~\ref{tab:finite-cubic-sic} shows results on $\textrm{SIC}(G)$ in finite cubic graphs, up to order $18$.

% {\color{blue}
% TODO:
% \begin{enumerate}
%      \item Determine SIC density for P2 cylinders with n = 2 or 4 mod 6 (and the solutions)     
%      \item Prove that $\textrm{SIC}(G_n) < n$ for $n \ne 6k$ or some ratio.
% \end{enumerate}
% \vspace{2em}
% }

\begin{table}[ht]
    \centering
    \begin{tabular}{c|r@{\hskip 1em}r@{\hskip 1em}r@{\hskip 1em}r@{\hskip 1em}r@{\hskip 1em}r@{\hskip 1em}r@{\hskip 1em}r@{\hskip 1em}r}
        $n$                           & 4 & 6 & 8 & 10 & 12 & 14  & 16 & 18 \\ \hline
        cubic graphs                           & 1 & 2 & 5 & 19 & 85 & 509 & 4060 & 41301 \\
        cubic graphs with SIC                      & 0 & 2 & 4 & 14 & 63 & 386 & 3189 & 33586  \\
        lowest value of $\textrm{SIC}(G)$      & - & 6 & 6 & 7  & 8  & 10  & 11   & 12          \\
        highest value of $\textrm{SIC}(G)$     & - & 6 & 7 & 9  & 12 & 12  & 15   & 18          \\

        %\hline
        % with RED:OLD                  & 1 & 1 & 4 & 15 & 67 & 409 & 3370 & 35323 & ?      \\
        % lowest $\textrm{RED:OLD}(G)$  & 4 & 6 & 6 & 7  & 8  & 10  & 11   & 12    & ?      \\
        % highest $\textrm{RED:OLD}(G)$ & 4 & 6 & 8 & 10 & 12 & 14  & 16   & 18  & ?

    \end{tabular}
    \caption{Results on SICs for finite (connected) cubic graphs}
    \label{tab:finite-cubic-sic}
\end{table}

\FloatBarrier
\subsection{SIC for Infinite Grids}\label{sec:grids}

In some of the following proofs, we will make use of what is known as a \emph{share argument} \cite{slat02a}.
For some detector vertex $v \in S$, the ``share'' of $v$ is a measure of how much $v$ contributes to the domination of its neighbors; more formally, $sh(v) = \sum_{u \in N[v]} \frac{1}{dom(u)}$, where $dom(u) = |N_S[u]|$ is the domination number of $u \in V(G)$.
If $L \in \mathbb{R}^+$ is an upper bound on the average share value of all detector vertices, then $L^{-1}$ is a lower bound on the density of $S$ in $G$.
We note that this technique is only valid (without other modifications) if $G$ exhibits the ``slow growth'' property mentioned in the introduction; all infinite graphs we consider in this paper have this required property.

\begin{figure}[ht]
    \centering
    \begin{tabular}{c@{\hskip 4em}c}
        \centered{\includegraphics[width=0.325\textwidth]{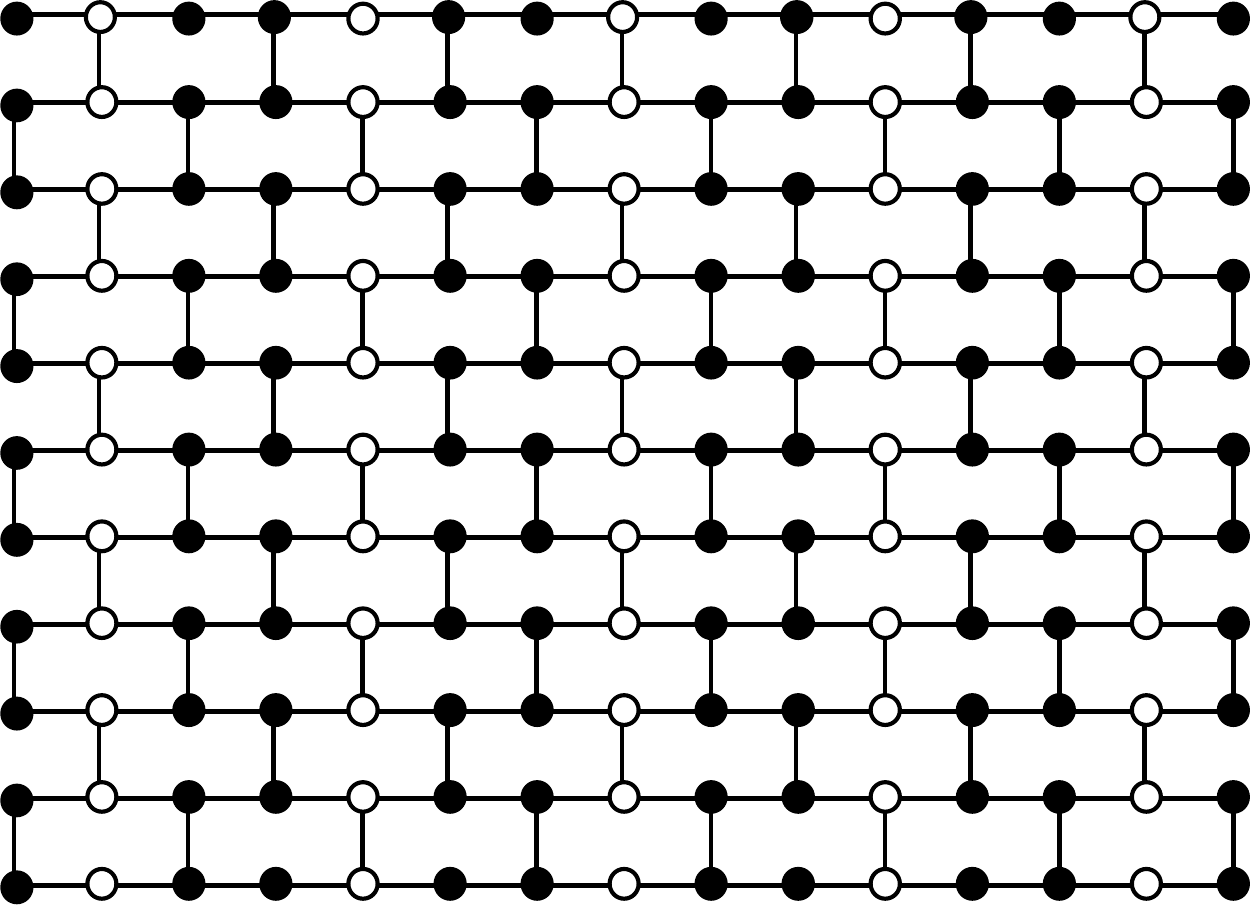}} & \centered{\includegraphics[width=0.325\textwidth]{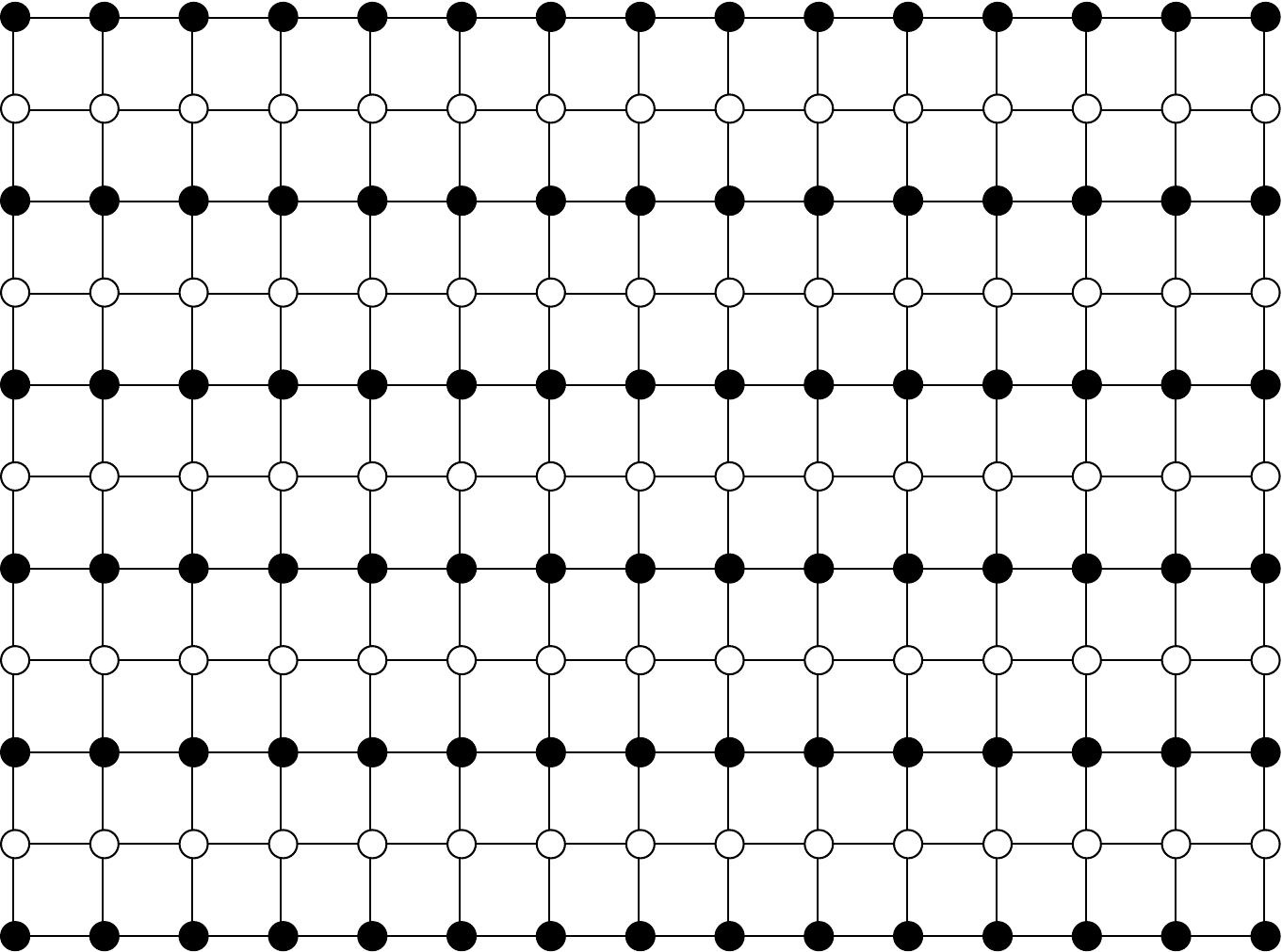}} \\
        (a) & (b) \\ \\
        \centered{\includegraphics[width=0.37\textwidth]{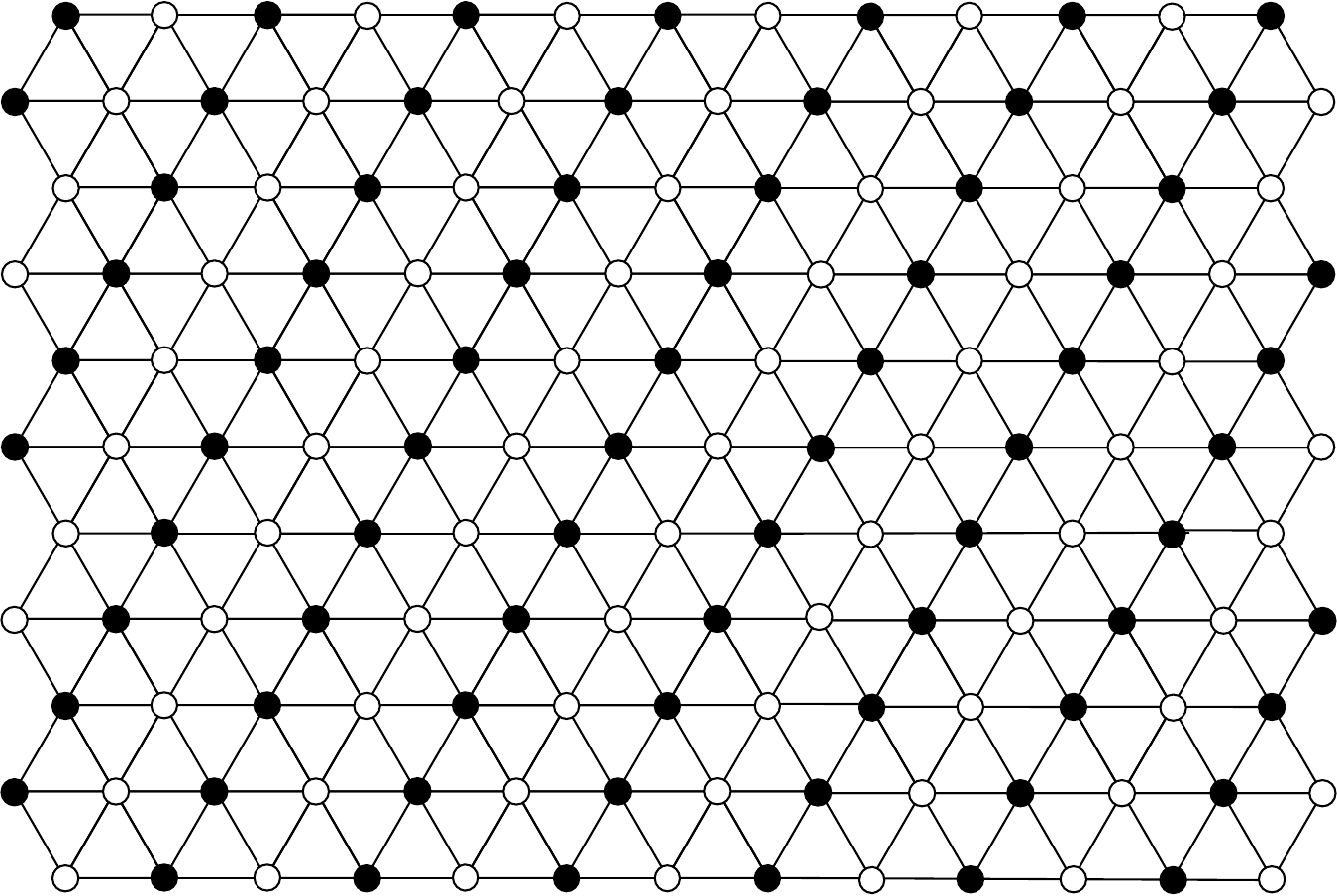}} & \centered{\includegraphics[width=0.325\textwidth]{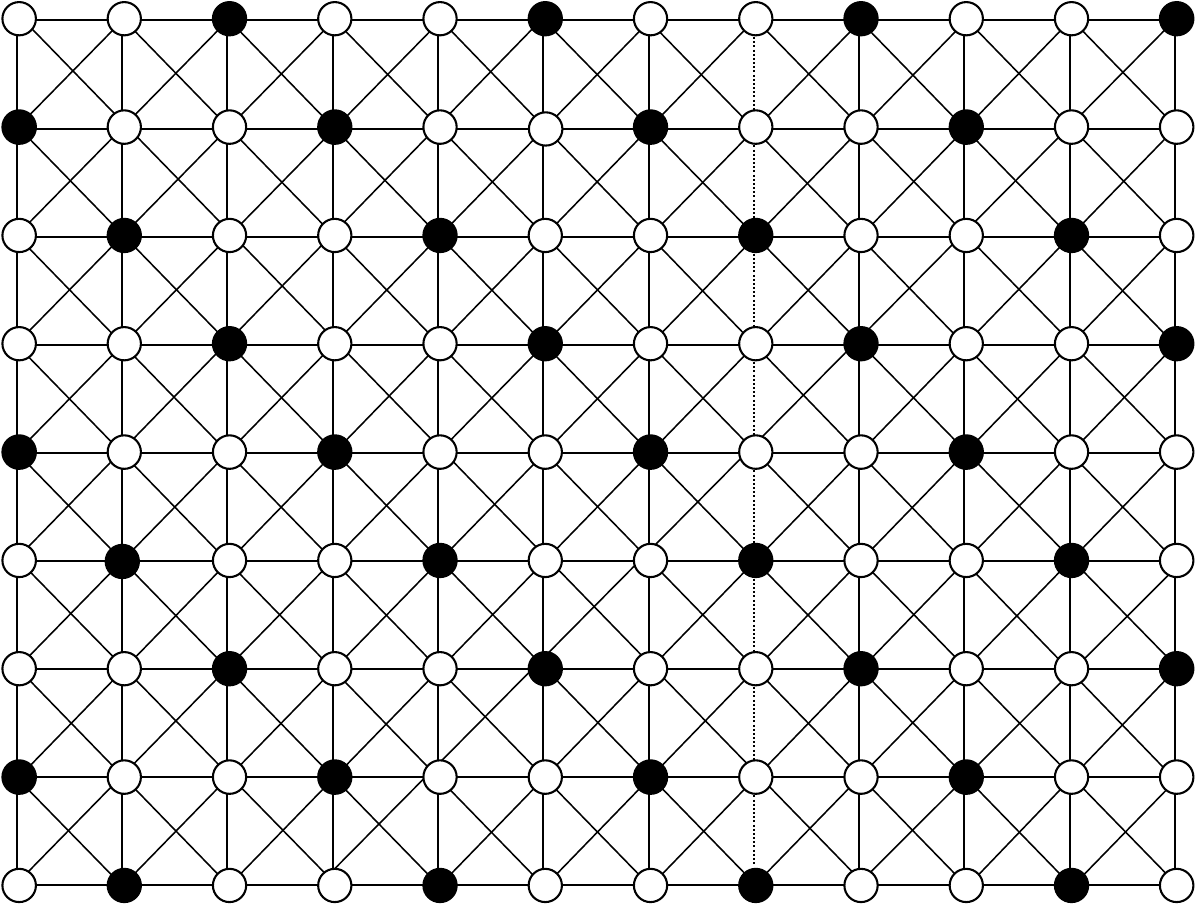}} \\
        (c) & (d) \\ \\
        \end{tabular}
    \caption{Our best constructions of SIC on HEX (a), SQR(b), TRI (c), and KNG (d). Shaded vertices denote detectors.}
    \label{fig:inf-grids-sic-solns}
\end{figure}

From Theorem~\ref{theo:sic-reg-lower-2-k}, we know that any 4-regular graph $G$ permitting SIC must have $\textrm{SIC}\%(G) \ge \frac{1}{2}$.
Figure~\ref{fig:inf-grids-sic-solns}~(b) gives an SIC for the infinite square grid with optimal density $\frac{1}{2}$.

Theorem~\ref{theo:sic-grid-values} presents tight bounds for SIC on several infinite grids; solutions achieving these optimal densities are presented in Figure~\ref{fig:inf-grids-sic-solns}.

\begin{theorem}\label{theo:sic-kng-tight}
The infinite king grid, KNG, has $\textrm{SIC\%}(G) = \frac{1}{3}$.
\end{theorem}
\begin{proof}
To demonstrate the upper bound, consider Figure~\ref{fig:inf-grids-sic-solns}~(d), which shows an SIC on KNG with density $\frac{1}{3}$.
The lower bound proof will proceed by demonstrating that any SIC set $S$ for KNG must be at least 3-dominating; that is, every vertex must be at least 3-dominated.
From Observation~\ref{obs:sic-no-leaves}, we know that any vertex $x \in S$ must be at least 3-dominated, so we need only show that any $x \notin S$ must likewise be at least 3-dominated.
Suppose to the contrary that some $x \notin S$ is only 2-dominated (the minumum domination number, as stated by Table~\ref{tab:ft-ic-cmp}).
We will use the notation that $\textrm{KNG} = (\mathbb{Z} \times \mathbb{Z}, \{ (v_{i,j}, v_{k,\ell}) : max(|i - k|, |j - \ell|) = 1\})$ and let $x = v_{0,0}$.
Because $x$ is exactly 2-dominated, we must place two distinct detector vertices $p,q \in S$ within $N(x)$.
If there exists some $r \in N(x)$ with $\{p,q\} \subseteq N(r)$, then $(x, r)$ cannot be distinguished, a contradiction.
Otherwise, without loss of generality, it must be that $p = v_{1,1}$ and $q = v_{\textrm{-}1,\textrm{-}1}$.
We then see that $(v_{1,0}, v_{2,0})$ are not distinguished, a contradiction, completing the proof.
\end{proof}

\begin{theorem}\label{theo:sic-k-a-b}
Let $G$ be a regular graph permitting some arbitrary domination-based parameter with uniform detection regions of size $k \ge 1$ and detector set $S \subseteq V(G)$.
If $a,b \ge 1$ are lower bounds for the average domination numbers of vertices in $S$ and $V(G)-S$, respectively, then a lower bound for the density of $S$ in $V(G)$ is given by $\frac{b}{k + b - a}$.
\end{theorem}
\begin{proof}
Let $c \in V(G)$ be any arbitrary vertex and let $x = \limsup_{r \to \infty} \frac{|B_r(c) \cap S|}{|B_r(c)|}$.
We know that
\begin{align*}
\limsup_{r \to \infty} \frac{\sum_{v \in B_r(c)} dom(v)}{|B_r(c)|} &= \limsup_{r \to \infty} \frac{k |B_r(c) \cap S|}{|B_r(c)|} \\
&= kx.
\end{align*}
And we know that
\begin{align*}
\limsup_{r \to \infty} \frac{\sum_{v \in B_r(c)} dom(v)}{|B_r(c)|} &\ge \limsup_{r \to \infty} \frac{a |B_r(c) \cap S|}{|B_r(c)|} + \limsup_{r \to \infty} \frac{b |B_r(c) - S|}{|B_r(c)|} \\
&= ax + b(1 - x).
\end{align*}
By combining these facts, $kx \ge ax + b(1 - x)$, implying that $x \ge \frac{b}{k + b - a}$, completing the proof.
\end{proof}

\begin{figure}[ht]
    \centering
    \includegraphics[width=0.25\textwidth]{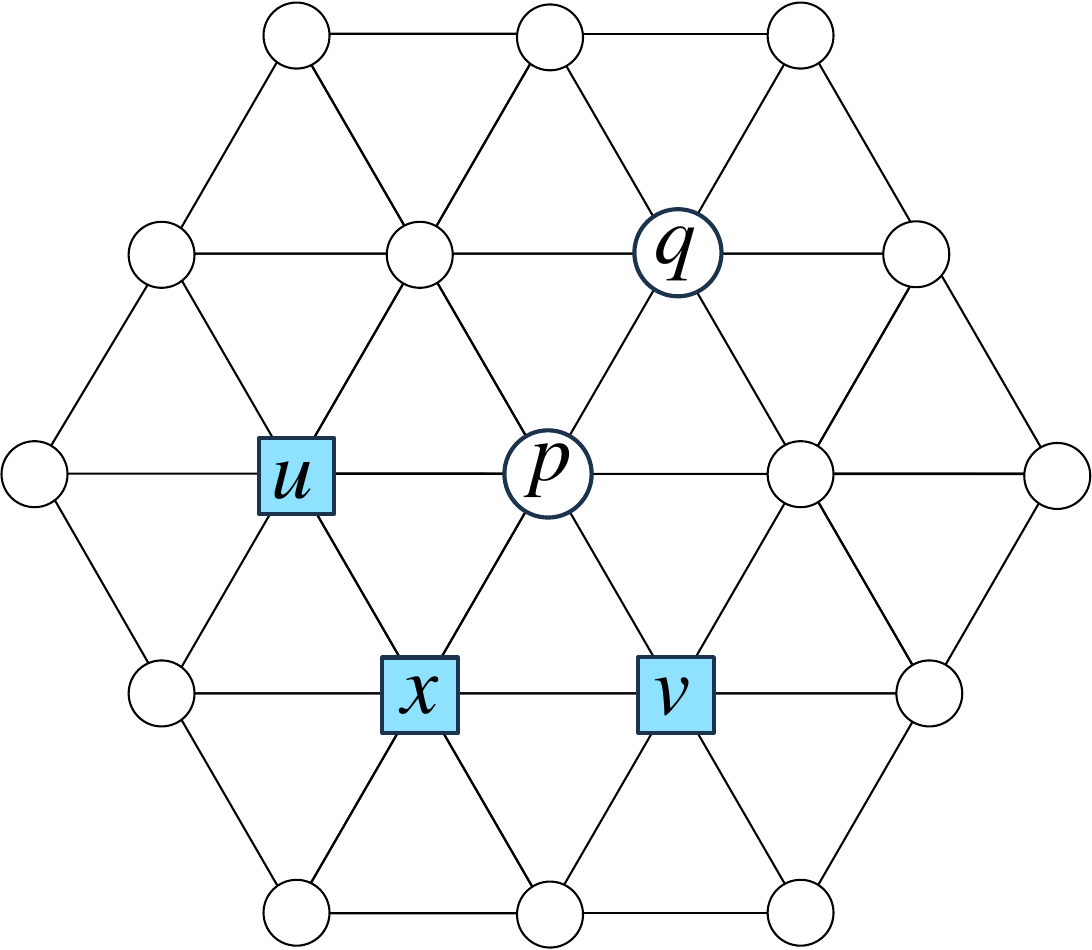}
    \caption{Forbidden non-detector configuration, $x, u, v$}
    \label{fig:sic-tri-tight-labeling}
\end{figure}

\begin{theorem}\label{theo:sic-tri-tight}
The infinite triangular grid, TRI, has $\textrm{SIC\%}(TRI) = \frac{1}{2}$.
\end{theorem}
\begin{proof}
To prove the upper bound, consider Figure~\ref{fig:inf-grids-sic-solns}~(c), which presents an SIC for TRI with density $\frac{1}{2}$.
To prove the lower bound, we will first show that any $x \notin S$ must have $dom(x) \ge 4$.
Let $x \notin S$, and consider the labeling shown in Figure~\ref{fig:sic-tri-tight-labeling}.
If $\{x,u,v\} \cap S = \varnothing$, then we see that $(p,q)$ cannot be distinguished, a contradiction.
Now, consider any non-detector $y \notin S$.
By the pigeonhole principle if we try to place three or more non-detectors in $N(y)$, then we will have a path of three non-detectors which is isomorphic to $uxv$, a contradiction.
Therefore, we can place at most two non-detectors in $N(y)$, implying $|N[y] \cap S| \ge 4$.
By substituting $a = 3$ and $b = 4$ into Theorem~\ref{theo:sic-k-a-b}, we arrive at a lower bound of $\frac{4}{7 + 4 - 3} = \frac{1}{2}$, completing the proof.
\end{proof}

% {\color{blue} check the TRI lower bound - did someone else prove this value? our lower bound is 3/7 < 1/2.}

\begin{theorem}\label{theo:sic-grid-values}
The upper and lower bounds for SIC:
\begin{enumerate}[label=\roman*]
    \item For the infinite hexagonal grid HEX, $\textrm{SIC\%}(HEX) = \frac{2}{3}$. 
    \item For the infinite square grid SQR, $\textrm{SIC\%}(SQR) = \frac{1}{2}$. 
    \item For the infinite triangular grid TRI, $\textrm{SIC\%}(TRI) = \frac{1}{2}$. 
    \item For the infinite king grid KNG,
$\textrm{SIC\%}(KNG) = \frac{1}{3}$.  
\end{enumerate}
\end{theorem}

\FloatBarrier
\bibliographystyle{ACM-Reference-Format}
\bibliography{refs, lob-refs}

\end{document}